\documentclass[aps,epsf,preprint]{revtex4-1}
\usepackage{amsmath,amssymb}

\def\1{{\bf 1}}
\def\[{\left[}
\def\]{\right]}
\def\be{\begin{eqnarray}}
\def\ee{\end{eqnarray}}
\def\bm{\begin{pmatrix}}
\def\em{\end{pmatrix}}
\def\nn{\nonumber}
\def\({\left(}
\def\){\right)}
\def\bk#1{\langle#1\rangle}
\def\eq#1{Eq.(\ref{#1})}

\def\a{\alpha}
\def\r{\rho}
\def\s{\sigma}
\def\o{\omega}
\def\e{\epsilon}

\def\f{\phi}

\def\l{\lambda}
\def\m{\mu}
\def\x{\times}

\def\p{\partial}
\def\d{\delta}
\def\l{\lambda}
\def\n{\nu}
\def\h{{1\over 2}}

\def\labels#1{\label{#1}}
\def\edc{\end{document}}

\def\bn{\begin{enumerate}}
\def\i{\item}
\def\en{\end{enumerate}}
\def\b{\beta}
\def\g{\gamma}

\def\ba{\begin{array}}
\def\ea{\end{array}}
\def\bc{\begin{center}}
\def\ec{\end{center}}

\def\edoc{\end{document}}

\def\^{$\wedge$}
\def\.{\!\cdot\!}

\def\igw#1{\includegraphics[width=#1cm]}
\def\+{\!+\!}
\def\-{\!-\!}
\def\={\!=\!}

\usepackage{color}

\def\M{M\"obius\ }
\def\Q{\Psi}
\def\pf{{\rm Pf}}

\def\igw#1{\includegraphics[width=#1cm]}
 
\usepackage{graphicx}
\graphicspath{/Users/harrylam/Current/filetex/Helicity/graphics/}

\begin{document}
\title{Off-Shell Yang-Mills Amplitude in the CHY Formalism}
\author{C.S. Lam$^{1,2,3}$}
\email{Lam@physics.mcgill.ca}
\address{$^1$Department of Physics, McGill University\\
 Montreal, Q.C., Canada H3A 2T8\\
$^2$Department of Physics and Astronomy, University of British Columbia,  Vancouver, BC, Canada V6T 1Z1 \\
$^3$CAS Key Laboratory of Theoretical Physics, Institute of Theoretical Physics, Chinese Academy of
Sciences, Beijing 100190, China\\}

\begin{abstract}
\noindent M\"obius invariance is used to construct gluon tree amplitudes in the Cachazo, He, and Yuan (CHY) formalism. If it is equally effective in steering the construction of  off-shell  tree
amplitudes, then the S-matrix CHY theory can be used to replace the Lagrangian Yang-Mills theory. 
In the process of investigating this possibility, we find that the CHY formula can indeed be modified
to obtain a M\"obius invariant off-shell amplitude, but unfortunately this modified amplitude $M_P$ is not the Yang-Mills amplitude because
it lacks gauge invariance. A complementary 
amplitude $M_Q$ must be added to restore gauge invariance, but its construction relies on the Lagrangian and 
not M\"obius invariance.  Although neither $M_P$
nor $M_Q$ is fully gauge invariant, both are partially gauge invariant in a sense to be explained. This
partial gauge invariance turns out to be  very useful for checking calculations. A Feynman amplitude
so split into the sum of $M_P$ and $M_Q$ also contains fewer terms.

\end{abstract}
\narrowtext
\maketitle
\section{Introduction}
S-matrix theory was popular in the 1960's, but it failed to take off because there was no 
way to incorporate interaction into it without a Lagrangian. This situation
changed in 2014 when Cachazo,
He, and Yuan (CHY) \cite{CHY1,CHY2,CHY3,CHY4,CHY5} came up with an S-matrix theory which can reproduce 
tree-level scattering of gluons, gravitons, and many others, with the additional advantage
that double-copy relations appear naturally. These refer to relations that are very difficult to understand
in the Lagrangian approach, linking together pairs of amplitudes such as graviton
amplitude  and the square of  Yang-Mills  amplitude.

$n$-body  CHY amplitudes are given by a complex
integral with  \M invariance, an invariance  crucial
in steering the construction of these amplitudes. Such construction enables  local interaction and local
propagation to appear  in an S-matrix theory,  a very remarkable feat
 because S-matrix a priori knows nothing about 
a local structure of space-time. This success raises the hope that maybe \M invariance is also
able to simulate fully  local space-time interaction, to reproduce off-shell tree amplitudes and 
hence loops without a Lagrangian. 

In the case of  $\f^3$ interaction, this is indeed the case. A simple modification of the scattering function enables all correct scalar Feynman tree diagrams to be reproduced, including those with off-shell external legs \cite{LY1, LY2}. In this article, we
examine to what extent \M invariance can also be used to reproduce off-shell Yang-Mills tree amplitudes.

In the case of off-shell Yang-Mills kinematics, \M invariance  forces not only a modification of the scattering function, as in the $\f^3$ case, but  also  a modification of the Pfaffian. This modified  $M_P$ describes an amplitude with a local interaction and  local propagation, but
unfortunately it  is not the correct Yang-Mills amplitude for $n>3$. The original on-shell $M_P$ is gauge invariant, but the modified off-shell $M_P$ retains only a partial gauge invariance. To restore full local gauge invariance, the hallmark of  the Yang-Mills theory, an additional term $M_Q$ must be added, which by itself also has partial but not full gauge invariance.

We will discuss how $M_Q$ can be obtained, but in its construction \M invariance is no longer a useful guide
when $n\ge 4$. Its appearance  is related  to the emergence of 
ghosts in Yang-Mills loops and off-shell
Yang-Mills tree amplitudes so it is unavoidable.

On-shell Yang-Mills amplitude in the CHY formalism is reviewed in Section II, to show the power of \M
invariance, and to see what modification is required to maintain the invariance for off-shell kinematics.
The details of such modifications will be discussed in Section III and Section IV. This modification
does enable $M_P$ to retain \M invariance off-shell, but an additional term $M_Q$ is needed to match
the Feynman amplitude $M_F$. 
 In Section V, we show  how $M_Q$ can be constructed, and illustrate the procedure with the
 explicit construction for $n=4$. The reason behind the necessary appearance of $M_Q$ 
 can be traced back to local gauge invariance, a topic which is discussed in Section VI.  Amplitudes for $n\ge 5$ are discussed in Sections VII, to illustrate how the Feynman amplitude can be simplified by its split into
 $M_P$ and $M_Q$, and to show how partial gauge invariance can be used to check calculations for a larger $n$. Section VIII provides a conclusion.

\section {\M invariant amplitude}
A color-stripped $n$-gluon scattering amplitude in the natural order $(12\cdots n)$ is given by the CHY formula
\cite{CHY2}  
\be
M_P&=&\(-{2g\over 2\pi i}\)^{n-3}\oint_\Gamma{\s_{(pqr)}^2\over\s_{(12\cdots n)}}\(\prod_{i=1,i\not=p,q,r}^n{d\s_i\over f_i}\)\ P,\labels{ms}\ee
where $g$ is the coupling constant  henceforth taken to be 1, 
$\s_{(pqr)}=\s_{pq}\s_{qr}\s_{rp},\
\s_{(12\cdots n)}=\prod_{i=1}^n\s_{i,i\+1}$ with $\s_{n\+1}\equiv\s_1$, and $\s_{ij}=\s_i\-\s_j$. 
The scattering functions $f_i$ are defined by
\be f_i&=&\sum_{j=1,j\not=i}^n{2a_{ij}\over \s_{ij}},\quad (1\le i\le n),\labels{f}\ee
with $k_i$ being the outgoing momentum of the $i$th gluon. The quantity $a_{ij}=a_{ji}$
is a linear function of scalar products of momenta whose explicit form will be discussed later.
The reduced Pfaffian $P=\pf'(\Q)$
is related to the Pfaffian of a matrix $\Q^{\l\n}_{\l\n}$ by
\be P=\pf'(\Q)={(-1)^{\l+\nu \+n\+1}\over\s_{\l\n}}\pf\(\Q^{\l\n}_{\l\n}\),\quad (\l<\n),\labels{ln}\ee
where $\Q^{\l\n}_{\l\n}$ is obtained from
 the matrix $\Q$ with its $\l$th and $\n$th columns and rows removed. 
The antisymmetric matrix $\Q$ is made up of three $n\x n$ matrices $A, B, C$,
\be
\Q=\bm A&-C^T\\ C&B\\ \em.\labels{QABCD}\ee
The non-diagonal elements of these three sub-matrices are
\be A_{ij}&=&{a_{ij}\over\s_{ij}},\quad B_{ij}={\e_i\.\e_j\over\s_{ij}}:={b_{ij}\over\s_{ij}},\nn\\
 C_{ij}&=&{c_{ij}\over\s_{ij}},\quad 
-C^T_{ij}={c_{ji}\over\s_{ij}},\quad 
(1\le i\not=j\le n),\labels{ABC}\ee
where $c_{ij}$ is a linear function of the scalar products $\e\.k$
whose exact form will be decided later,  and  $\e_i$ is the polarization of the $i$th gluon.
The diagonal elements of $A$ and $B$ are zero, and that of $C$ is defined by
\be C_{ii}=-\sum_{j=1}^nC_{ij},\labels{CC}\ee
so that $\sum_jC_{ij}=0$ for all $i$. A similar property is true for $A$ if the scattering equations $f_i=0$
are obeyed. This is the case because the integration contour $\Gamma$ encloses these zeros anticlockwise. 

The factors in \eq{ms} are designed to transform covariantly under the \M transformation
\be \s_i\to {\a\s_i+\b\over\g\s_i+\d},\quad (\a\d-\b\g=1),\labels{mt}\ee
in such a way that the total weight of the integrand is zero, thus resulting in a \M invariant integrand. 
Specifically, under the \M transformation, if we let $\l_i=1/(\g\s_i+\d)$, then 
\be
d\s_i&\to&\l_i^2d\s_i,\nn\\
\s_{ij}&\to&\l_i\l_j\s_{ij},\nn\\
\s_{(p,q,r)}&\to&(\l_p\l_q\l_r)^2\s_{(p,q,r)},\nn\\
\s_{(12\cdots n)}&\to&\(\prod_{i=1}^n\l_i^2\)\s_{(12\cdots n)}.\labels{mt1}\ee
The scattering function transform covariantly like
\be
f_i&\to&\l_i^{-2}f_i,\labels{mtf}\ee
as long as
\be
\sum_{j=1,j\not=i}^na_{ij}=0.\labels{conda}\ee
Thus the integrand of \eq{ms} is \M invariant as long as
\be
P&\to&\(\prod_{i=1}^n\l_i^{-2}\)P\labels{mtP}\ee
whatever $p,q,r$ are.

Using \eq{mt1}, as well as \eq{QABCD} to \eq{CC}, we see that 
$P=\pf'(\Psi)$ in \eq{ln} does transform that way, whatever $\l,\n$ are, provided
\be
C_{ii}&\to&\l_i^{-2}C_{ii},\labels{mtC}\ee
which is the case if
\be\sum_{j=1,j\not=i}^nc_{ij}=0.\labels{condc}\ee

As long as \eq{ms} is \M invariant, the integral $M_P$ can be shown to be independent of the choice
of $p,q,r$, as well as the choice of $\l,\n$. To be invariant, $a_{ij}$ and $c_{ij}$ must be chosen to satisfy
\eq{conda} and \eq{condc}.

For on-shell gluons with transverse polarization,
$k_i^2=0$ and $\e_i\.k_i=0$, momentum conservation guarantees these conditions to be 
satisfied if
\be
a_{ij}&=&k_i\.k_j:=a'_{ij},\nn\\
c_{ij}&=&\e_i\.k_j:=c'_{ij},\labels{acp}\ee
which is the choice in the CHY theory.
For off-shell kinematics with possibly longitudinal and time-like polarizations, $k_i^2\not=0$
and $\e_i\.k_i:=d_i\not=0$, \eq{acp} no longer satisfies \eq{conda} and 
\eq{condc}, so the expression for $a_{ij}$ and $c_{ij}$ must be modified. How this can be done will be discussed in the next two sections.

\section{$\boldsymbol{a_{ij}}$ determined by the propagators}
Let
\be
a_{ij}&=&a'_{ij}+\r_{ij},\nn\\
c_{ij}&=&c'_{ij}+\eta_{ij}.\labels{ac}\ee
The constraints \eq{conda} and \eq{condc} restrict the additional terms to satisfy
\be
\sum_{j\not=i,j=1}^n\r_{ij}&=&k_i^2,\labels{rho}\\
\sum_{j\not=i,j=1}^n\eta_{ij}&=&\e_i\.k_i:=d_i.\labels{xi}\ee
In this section we will discuss how to obtain $\r_{ij}=\r_{ji}$, leaving the determination of $\eta_{ij}$ to the next section.

\eq{rho} alone is not sufficient to determine all $\r_{ij}$. Since we want to retain local propagation for off-shell amplitudes, we demand \eq{ms} to yield  correct propagators in the Feynman gauge. 
For the color-stripped amplitude $M_P$ in natural order,
this requires
$\sum_{i\not=j;i,j\in{\cal D}}a_{ij}=\(\sum_{i\in{\cal D} }k_i\)^2:=s_{\cal D}$ for every consecutive set of numbers
${\cal D}$. This requirement has a unique solution for $\r$ given by \cite{LY1,LY2}
\be \r_{i,i\pm 1}&=&+\h(k_i^2+k_{i\pm 1}^2),\nn\\
\r_{i\mp 1,i\pm 1}&=&-\h k_i^2,\nn\\
\r_{ij}&=&0\quad{\rm otherwise}.\labels{rhosol}\ee
where all indices are understood to be mod $n$. 

There is another way to retain \M covariance of $f_i$ off-shell without modifying $a_{ij}=a'_{ij}$:
one can add  an extra dimension  and use the extra momentum component to simulate $k_i^2$. However,
this does not retain local
propagation as the resulting propagators turn out to be incorrect.

\section{$\boldsymbol{c_{ij}}$ determined by the triple-gluon vertex}
There are also many solutions of $\eta_{ij}$ to satisfy \eq{xi}, but unlike
$\r_{ij}$ which can be fixed by the local propagation requirement, there is no obvious way
to settle what $\eta_{ij}$ should be.

One of the many solutions of  \eq{xi} is
\be
c_{i,i\pm 1}&=&c'_{i,i\pm 1}+\h d_i,\nn\\
c_{ij}&=&c'_{ij}\quad{\rm otherwise}.\labels{xisol}\ee
We shall adopt this solution throughout because it is the simplest and because it yields the correct $n=3$ off-shell amplitude.

To see that, recall that the triple-gluon vertex (with a unit coupling constant, and the color factor stripped) 
depicted in Fig.1 is
\be V&=&\e_1\.\e_2\ \e_3\.(k_1\-k_2)+\e_2\.\e_3\ \e_1\.(k_2-k_3)+\e_3\.\e_1\ \e_2\.(k_3-k_1)\nn\\
&=&b_{12}(c'_{31}\-c'_{32})+b_{23}(c'_{12}\-c'_{13})+b_{31}(c'_{23}\-c'_{21}).\labels{V}\ee
Using \eq{xisol}, this becomes
\be
V=b_{12}(c_{31}\-c_{32})+b_{23}(c_{12}\-c_{13})+b_{31}(c_{23}\-c_{21})
=2(\-b_{12}c_{32}\+b_{23}c_{12}\-b_{31}c_{21}),\labels{VV}\ee
which is precisely what \eq{ms} yields when $n=3$.  Therefore, the choice of \eq{xisol} enables
the triple-gluon vertex to be reproduced correctly by $M_P$ in \eq{ms} for $n=3$.

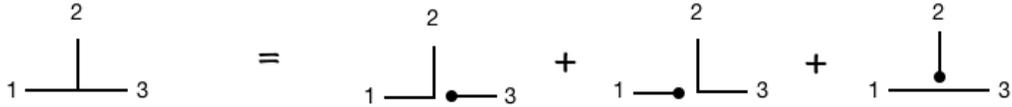
\bc\igw{14}{Fig1}\\ Fig.1\quad Triple gluon vertex and its three sub-diagrams\ec

It is convenient to represent each of the three terms in \eq{V} by a separate sub-diagram,
as shown on the right  of Fig.1. This pictorial representation makes it easier to distinguish
different terms in a Feynman
diagram. 

The reason to use \eq{xisol} also for $n>3$ is the following. It turns out that no matter how $\eta_{ij}$
is chosen, there is no way  to convert all $c'_{ij}$ into $c_{ij}$ when $n>3$, thereby enabling $M_P$ to be
the off-shell Feynman amplitude. For that reason any choice of $\eta_{ij}$ is equally good, so we 
might as well use \eq{xisol}, which not only reproduces the triple-gluon vertex, but is also the simplest
solution of \eq{xi}.

To show that there is no way to convert all $c'_{ij}$ into $c_{ij}$,  consider $n=4$. 
There are many Feynman sub-diagrams but let us just look at the four
shown in Fig.2.  

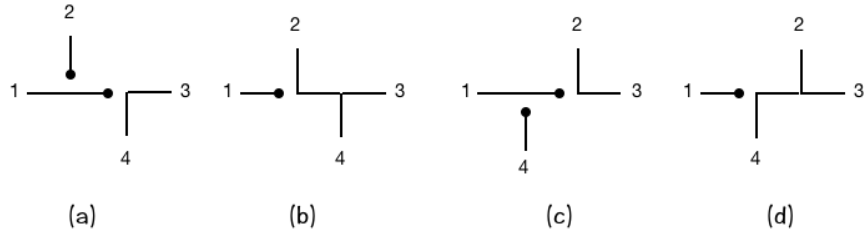
\bc\igw{12}{Fig2}\\Fig.2\quad Four $n=4$ Feynman sub-diagrams\ec

 All four contain a factor involving some combination of $c'_{1j}$. That factor is
$c'_{13}-c'_{14}$ in Fig.2(a), $c'_{12}-(c'_{13}+c'_{14})$ in Fig.2(b), $c'_{12}-c'_{13}$ in Fig.2(c),
and $(c'_{12}+c'_{13})-c'_{14}$ in Fig.2(d). To convert all these combinations of $c'$ into the corresponding
combinations of $c$, we must require
\be
\eta_{13}-\eta_{14}&=&0,\nn\\
\eta_{12}-(\eta_{13}+\eta_{14})&=&0,\nn\\
\eta_{12}-\eta_{13}&=&0,\nn\\
(\eta_{12}+\eta_{13})-\eta_{14}&=&0.\labels{cpc}\ee
Moreover, \eq{xi} also requires $\eta_{12}+\eta_{13}+\eta_{14}=d_1$. There are just too many equations
for $\eta_{1j}$ to have a solution. Thus it is not possible to convert all the $c'_{ij}$ appearing in all the $n=4$ Feynman diagrams into $c_{ij}$, no matter now $\eta_{ij}$ are chosen. For a larger $n$, it is even worse because there will be more equations to satisfy.

$M_P$ in \eq{ms} contains only $a_{ij}, b_{ij}, c_{ij}$, but no $d_i$, it clearly cannot be equal
to the Feynman amplitude $M_F$ for Yang-Mills theory which is a function of $a'_{ij}, b_{ij}, c'_{ij}$,
unless all $a'$ and $c'$ can be converted into $a$ and $c$ without the appearance of $k_i^2$ and $d_i$.
Since this is impossible for $n\ge 4$, an additional term
$M_Q=M_F-M_P$
must be present.

\section{ Method to compute $\boldsymbol{M_Q}$ illustrated with $\boldsymbol{n=4}$}
$M_Q=M_F(a',b,c')-M_P(a,b,c)$ can be obtained by using Feynman rules to compute $M_F$, and \eq{ms}
to compute $M_P$.  Since there are many terms in $M_F$ and many terms
in $M_P$, this computation turns out to be quite tedious even for $n=4$. It is much worse for larger $n$.

Fortunately, with the following observation there is a much simpler way to compute $M_Q$. For on-shell gluons
with transverse polarization, where $a=a'$ and $c=c'$, we know that $M_P$ gives the correct Yang-Mills amplitude,
\be
M_F(a',b,c')=M_P(a',b,c').\labels{FP}\ee
For off-shell kinematics, the Feynman rules remain the same, so $M_F$ is not changed. If we use 
\eq{ac} to convert $a'$ and $c'$ in $M_F$ into $a$ and $c$, then \eq{FP} implies that those terms
without the presence of any off-shell parameter $k_i^2, d_i$ must add up to give $M_P(a,b,c)$. The
remaining terms which contain at least one off-shell parameter must add up to give $M_Q$. Thus
$M_Q$ can be computed just by extracting those terms in $M_F$ that contain off-shell parameters.

Let us illustrate how to do that for $n=4$. The Feynman amplitude $M_F$ has an $s$-channel diagram
with 9 terms, a $t$ channel diagrams with 9 terms, and a four-gluon diagram with 3 terms. The four-gluon
terms consist of products $b_{ij}b_{kl}$, where $(ijkl)$ is a permutation of  (1234). Since neither $a'$ nor $c'$
enters, it cannot contribute to $M_Q$, so we will ignore it from now on.

The 18 $s$-channel and $t$-channel sub-diagrams are given in Fig.3. 

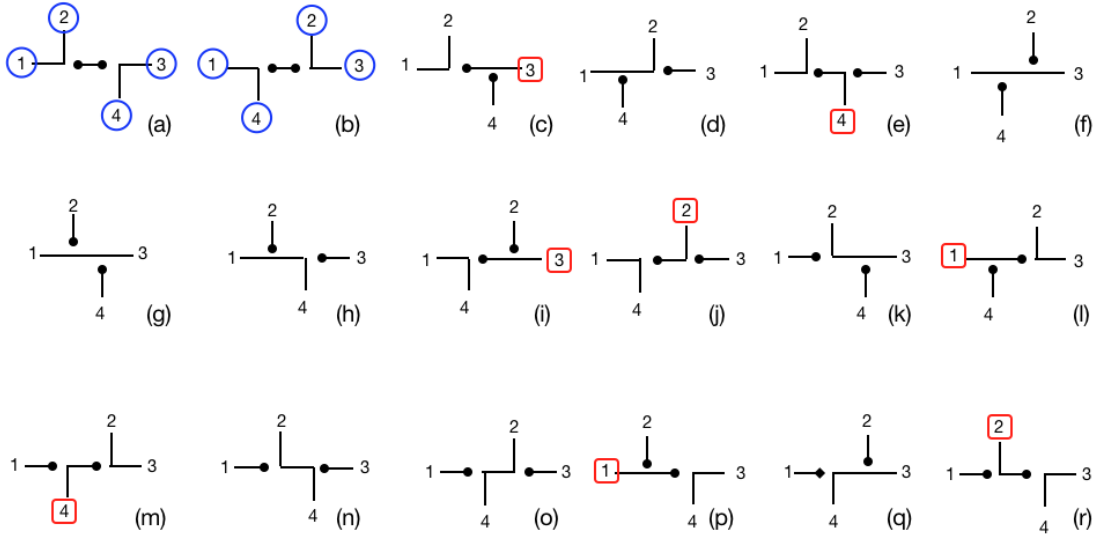
\bc\igw{15}{Fig3}\\ Fig.3\quad The 18 $s$ and $t$ channel Feynman sub-diagrams for $n=4$. Line numbers
enclosed by a box contributes to $d_i$, and line numbers enclosed by a circle contributes to $k_i^2$ in
$M_Q$\ec

Using the recipe given above, 
$M_Q$  turns out to be
\be
M_Q
&=&\(\sum_{i=1}^4k_i^2\)\({b_{12}b_{34}\over s}\+{b_{41}b_{23}\over t}\)
-\[
{b_{12}\over s}\(d_3c_{43}+d_4c_{34}\)+{b_{41}\over t}\(d_2c_{32}+d_3c_{23}\)\right.\nn\\
&+&\left.{b_{23}\over t}(d_1c_{41}+d_4c_{14})+{b_{34}\over s}\(d_1c_{21}\+d_2c_{12}\)\],\labels{mq4}\ee
where $s=s_{12}=(k_1\+k_2)^2=s_{34}=(k_3\+k_4)^2$ and $t=s_{41}=(k_4+k_1)^2=s_{23}=(k_2+k_3)^2$.

Note that there are ten terms in \eq{mq4} but 18 diagrams in Fig.3, so some of those diagrams must not contribute to $M_Q$.
To identify the diagrams that do not contribute to $M_Q$, 
let us first recall the meaning of the graphical components in sub-diagrams. 
A line ending with a heavy dot (which we  shall refer to as a `hammer') represents $c'_{il}-c'_{ir}$, with $i$
on the handle and $l,r$ to the left and right of the hammer head (the heavy dot). If $k_l$ or $k_r$ is an internal momentum,
it must be converted into the appropriate sum of external momenta, and $c'_{il}, c'_{ir}$ are then the corresponding sum of $c'$ between $i$ and these external momenta. With 
a similar notation, a heavy dot at both ends of a line (which we  shall call a `dumbbell') represents the factor $a'_{l_1l_2}-a'_{l_1r_2}-a'_{r_1l_1}+a'_{r_1r_2}$, where $l_i, r_i$ represent the lines to the left and to the right
of the two dumbbells (heavy dots) $i=1,2$. 

Graphically, the conversion equations
\eq{ac}, \eq{rhosol},  \eq{xisol} say that $d_i/2$ appears at a hammer handle  either when one and only one of
its two neighboring lines appears in the hammer strike region, or, when both appear, they appear on the same side of the hammer head. For example, there are two hammers in Fig.3(c), one at line 3 and one at line 4. The neighboring lines of 3 are 4 and 2, only one of them appears in the hammer strike region of 3, so $d_3$ appears. This is indicated in the diagram with a box around the number 3. The neighboring lines of 4 are 2 and 3, they appear in the hammer strike region of 4 on different sides, so $d_4$ does not enter, which is indicated in the diagram by the absence of a square box around the number 4. The emergence of $k_i^2$ in the dumbbell region,
indicated by a circle around the line number, can be obtained similarly.

In this way we can see where $d_i$ and $k_i^2$ appear in all  the diagrams in Fig.3. 
In particular, no $d_i$ is present in Figs.~3(d), (f), (g), (h), (k), (n), (o), and (q),
so these diagrams do not contribute to $M_Q$.
The 8 $d_i$ terms in \eq{mq4} come respectively from diagrams (c), (e), (j), (i), (l), (m), (p), and (r). Similar considerations applied to the dumbbell regions tell us where to put a circle to indicate the appearance of $k_i^2$.

Note that $b_{13}$ comes from diagrams 3(f) and 3(g) and $b_{24}$ comes from diagrams 3(n) and 3(o). The absence of these diagrams in $M_Q$ is the reason why neither $b_{13}$ nor $b_{24}$ appears in \eq{mq4}.

Note also that $M_Q$ is invariant under cyclic permutation. This should be the case because both $M_F$ and $M_P$ are invariant. When we permute \eq{mq4} from  (1234) to (2341), we get, for example,
\be {b_{12}b_{34}\over s_{12}}&\leftrightarrow& {b_{23}b_{41}\over s_{23}},\nn\\
{b_{12}\over s_{12}}(d_3c_{43}+d_4c_{34})&\to&{b_{23}\over s_{23}}(d_4c_{14}+d_1c_{41}),\nn\\
{b_{41}\over s_{41}}\(d_2c_{32}+d_3c_{23}\)&\to&{b_{12}\over s_{12}}\(d_3c_{43}+d_4c_{34}\),\quad etc.,
\nn\ee
showing explicitly that \eq{mq4} is cyclic permutation invariant.

It is  amusing to find out whether $M_Q$ can be written in the form of \eq{ms}. Namely, whether
there exists a \M covariant function $Q=Q(A_{ij},B_{ij},C_{ij},d_i,k_i^2)$ which transforms with a weight factor $(\l_1\l_2\l_3\l_4)^{-2}$, such that
\be
M_Q&=&\(-{2g\over 2\pi i}\)^{n-3}\oint_\Gamma{\s_{(pqr)}^2\over\s_{(12\cdots n)}}\(\prod_{i=1,i\not=p,q,r}^n{d\s_i\over f_i}\)\ Q.\labels{msq}\ee
Since the dependence of $M_Q$ on $a,b,c$ is assumed to arise from the dependence of $Q$ on 
$A,B,C$, it is clear from \eq{mq4} that if such a $Q$ exists, it must be 
\be
Q
&=&\[\(\sum_{i=1}^4k_i^2\)\({b_{12}b_{34}\over \s_{12}\s_{34}}\+{b_{14}b_{23}\over \s_{14}\s_{23}}\)-
{b_{12}\over \s_{12}}\(-d_3{c_{43}\over\s_{43}}+d_4{c_{34}\over\s_{34}}\)-{b_{14}\over \s_{14}}\(-d_2{c_{32}\over\s_{32}}+d_3{c_{23}\over\s_{23}}\)\right.\nn\\
&&\left.\ \ -{b_{23}\over \s_{23}}\(-d_4{c_{14}\over\s_{14}}+d_1{c_{41}\over\s_{41}}\)-{b_{34}\over \s_{34}}\(-d_1{c_{21}\over\s_{21}}+d_2{c_{12}\over\s_{12}}\)\]{1\over\s_{31}\s_{24}}.\labels{Q4}\ee
The extra factor $1/\s_{31}\s_{24}$ outside of the square brackets is there to enable $Q$ to transform with the correct covariant weight, and 
the signs of the various terms are needed to ensure  $M_Q$ to be reproduced after the $\s$-integrations.
With this $Q$, it turns out that $M_Q$ computed using \eq{msq} is indeed the correct $M_Q$ given by \eq{mq4}.

Although $Q$ exists for $n=4$, \M invariance cannot determine its form, nor that of $M_Q$,
so its existence is merely of academic interest. Unlike
 $P$, where \M invariance, permutation 
symmetry, and dimensional analysis largely determine what it should be, nothing similar is available
for $Q$. For example, without the Feynman diagrams and the discussion earlier in this section, there is no way even
to know that neither $B_{13}$ nor $B_{24}$ is present in $Q$. For that reason we shall no longer discuss
$Q$ from now on.

\section{ Local gauge invariance}

\subsection{Slavnov-Taylor identity}
The emergence of $M_Q$ can be traced back to local gauge invariance, the hallmark of Yang-Mills theory.
An amplitude possessing local gauge invariance must satisfy the 
Slavnov-Taylor identity \cite{S,T}, which relates
the divergence of an $n$-gluon Green's function  to the Green's function with $(n\-2)$ gluons and a ghost anti-ghost pair:
\be -{\p\over\p x_i^{\m_i}}\bk{A_{\m_1}^{a_1}(x_1)A_{\m_2}^{a_2}(x_2)\cdots A_{\m_n}^{a_n}(x_n)}=\sum_{k\not=i}
\bk{\bar\o^{a_i}(x_i)A_{\m_2}^{a_2}(x_2)\cdots D_{\m_k}\o^{a_k}(x_k)  \cdots A_{\m_n}^{a_n}(x_n)}.\labels{STI}\ee
 $A$ is the gluon field, $\o, \bar\o$ are the ghost and anti-ghost fields, and
$(D_\m\o)^a=\p_\m\o^a+gf_{abc}A_\m^b\o^c$ is the covariant derivative of the ghost field.
The corresponding relation for color-stripped amplitudes is shown in Fig.4, where solid lines are
gluons and dotted lines are ghosts. A cross ($\x$) at line $j$ represents the factor $d_j=\e_j\.k_j$, and a box ( \raisebox{-.3mm}{$\blacksquare$}\ ) at line $j$ represents the factor $k_j^2$. The cross comes from the derivative
of the ghost field, and the box is there to amputate the external leg in the $A\o$ term of $D\o$.

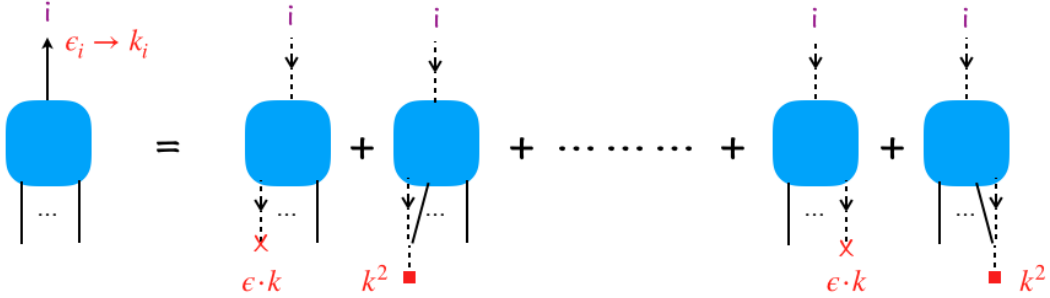
\bc\igw{15}{Fig4}\\ Fig.4\quad The Slavnov-Taylor relation relating the divergence of a gluon amplitude
to the covariant derivative on the ghost lines of gluon-ghost amplitudes\ec
 In tree order, this relation
can be derived directly from the gluon tree amplitude by replacing
$\e_i$ in a gluon line by $k_i$ \cite{FL}. 
Let us illustrate how that is done for $n=3$ and $i=2$.

Using the notation $\d_i({\cal O})$ to indicate replacing $\e_i$ in ${\cal O}$ by $k_i$, we get from \eq{V} that
\be
\d_2(V)&=&\e_1\.k_2\ \e_3\.(k_1-k_2)+k_2\.\e_3\ \e_1\.(k_2\-k_3)+\e_3\.\e_1\ k_2\.(k_3\-k_1)\nn\\
&=&-\e_1\.k_1\ \e_3\.k_1+\e_1\.k_3\ \e_3\.k_3+k_1^2\ \e_1\.\e_3-k_3^2\ \e_1\.\e_3,
\labels{ST3}\ee
where momentum conservation has been used to obtain the second line.
These four terms are depicted by the four diagrams in Fig.5, where 5(a), 5(b) correspond to
the first diagram on the right of Fig.4, respectively for $j=1$ and $j=3$, and 5(c), 5(d) correspond to the
second diagram. The $\e_3\.k_1$ factor in the first term comes from the gluon-ghost vertex in 5(a). 
The minus signs came from color ordering before color is stripped. 

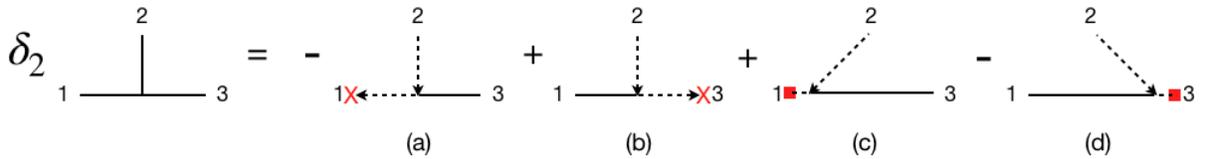
\bc\igw{16}{Fig5}\\ Fig.5\quad The Slavnov-Taylor identity for $n=3$\ec

What is important for our subsequent discussion is that $\d_i(M)$ for a local gauge invariant amplitude $M$
consists of terms proportional to $d_j$ and $k_j^2$ for all $j\not=i$, but it does not contain terms involving
$k_i^2$ in leading order of the off-shell parameters.  We shall refer to this absence of
$k_i^2$ as {\it partial gauge invariance}. It turns out that neither $M_P$ nor $M_Q$ is locally gauge
invariant, though their sum is, but both have partial gauge invariance. This  property is useful in checking the calculations of $M_P$ and $M_Q$, and puts a constraint on the allowed forms of $M_P$ and $M_Q$.

\subsection{$\boldsymbol{M_P}$ does not have local gauge invariance but it is partially gauge invariant}
Let us compute $\d_2(\Psi^{13}_{13})$ to see whether $\d_2(M_P)$ satisfies the Slavnov-Taylor identity.
The change $\e_2\to k_2$ leads to  $c'_{2j}\to a'_{2j},\ b_{2j}=b_{j2}\to c'_{j2}$, which in turn
leads to a change of $\Psi^{13}_{13}$ in 
the ($n$th) row and column containing $C_{2j}$ and $B_{2j}$. These changes are given by
\be
\d_2 d_2&=&k_2^2,\nn\\
\d_2b_{2j}&=&\d_2b_{j2}=c'_{j2}=c_{j2}-\h d_j, \quad(j=1,3)\nn\\
\d_2b_{2j}&=&\d_2b_{j2}=c'_{j2}=c_{j2},\quad (j\not=1,2,3),\nn\\
\d_2c_{2j}&=&\d_2c'_{2j}+\h d_2= a'_{2j}+\h d_2=a_{2j}-\h k_j^2,\quad(j=1,3)\nn\\
\d_2c_{24}&=&\d_2c'_{24}= a'_{24}=a_{24}+\h k_3^2,\nn\\
\d_2c_{2n}&=&c'_{2n}= a'_{2n}=a_{2n}+\h k_1^2,\nn\\
\d_2c_{2j}&=&\d_2c'_{2j}= a'_{2j}=a_{2j},\quad(j\not=1,2,3,4,n).\labels{d2}\ee
All other elements of $b_{ij}, c_{ij}, d_i$, and all elements of $a_{ij}$ remain the same. 

 We shall compute $\d_2(M_P)$ using the property that subtracting the $n$th row/column
from the first row/column of $\d_2\(\Psi^{13}_{13}\)$ does not change its Pfaffian.
The first row of $\Psi^{13}_{13}$ consists of 
\be \(0, A_{24},A_{25},\cdots,A_{2,n\-1},A_{2n},-C_{12},-C_{22},-C_{32},-C_{42},\cdots,-C_{n2}\),\nn\ee 
none of which is affected by $\d_2$ except
$-C_{22}$, 
\be
-\d_2C_{22}=\sum_{j\not=2}{\d_2c_{2j}\over\s_{2j}}=\(\sum_{j\not=2}A_{2j}\)-\h k_1^2\({1\over\s_{21}}-{1\over\s_{2n}}\)-\h k_3^2\({1\over\s_{23}}-{1\over\s_{24}}\) .\labels{c22}\ee
The $n$th row of $\Psi^{13}_{13}$ consists of 
\be\(C_{22},C_{24},C_{25},\cdots,C_{2,n\-1},C_{2n},B_{21},0,B_{23},B_{24},\cdots,B_{2n}\),\nn\ee
which under $\d_2$ is changed into
\be
&&\(\d_2C_{22},{\d_2c_{24}\over\s_{24}},{\d_2c_{25}\over\s_{25}},\cdots,{\d_2c_{2,n\-1}\over\s_{2,n\-1}},{\d_2 c_{2n}\over\s_{2n}},{\d_2b_{21}\over\s_{21}},0,{\d_2b_{23}\over\s_{23}},{\d_2b_{24}\over\s_{24}},\cdots,
{\d_2b_{2n}\over\s_{2n}}\)\nn\\
&=&\(\d_2C_{22},\hat A_{24},A_{25},\cdots,A_{2,n\-1},\hat A_{2n},0,
-\hat C_{12},0,-\hat C_{32},-C_{42},\cdots,-C_{n2}\),\labels{ppn}\ee
where
\be
\hat A_{24}&=&A_{24}\+\h {k_3^2\over\s_{24}},\nn\\
\hat A_{2n}&=&A_{2n}\+\h {k_1^2\over\s_{2n}},\nn\\
\hat C_{12}&=&C_{12}\-\h{d_1\over\s_{12}},\nn\\
\hat C_{32}&=&C_{32}\-\h{d_3\over\s_{32}}.\ee
Subtracting the $n$th row/column from the first row/column changes the first row into
\be
-\h\(0,{k_3^2\over\s_{24}},0,\cdots,0,{k_1^2\over\s_{2n}},{d_1\over\s_{21}},2\d_2C_{22},{d_3\over\s_{23}},0,\cdots,0\),\ee
and the first column into the same thing with a minus sign, leaving the rest of $\d_2(\Psi^{13}_{13})$ unchanged. The modified matrix contains only off-shell parameters $d_j, k_j^2$  in the first row/column,
so every term in $\pf\(\d_2\(\Q^{13}_{13}\)\)$, and thus every term in $\d_2(M_P)$, must be proportional to an off-shell parameter. Thus
\bn
\i  $\d_2(M_P)=0$ for on-shell gluons
with transverse polarization, as we already know;
\i $k_2^2$ and all $d_j, k_j^2$ for $j\ge 4$ are missing from $\d_2(M_P)$, hence $M_P$ cannot satisfy
the Slavnov-Taylor identity in which all $k_j^2$ and $d_j$ for $j\not=2$ must be present. This is
why  $M_Q$ is needed to restore local gauge invariance of the amplitude;
\i $M_P$ is invariant under permutation of the particles, thus if $k_2^2$ is absent from $\d_2(M_P)$, $k_i^2$
must be absent from $\d_i(M_P)$. By definition, $M_P$ has  partial gauge invariance; 
\i since both $M_F$ and $M_P$ have partial gauge invariance, $M_Q$ must also have partial gauge invariance.
\en 

\subsection{Partial gauge invariance of ${\boldsymbol{M_Q}}$ for ${\boldsymbol{n=4}}$}
Partial gauge invariance is a useful tool for verifying calculations. Together with cyclic permutation invariance,
it provides a non-trivial constraint on the allowed forms of $M_Q$. Let us illustrate these points with $n=4$.

For convenience, \eq{mq4} of $M_Q$ for $n=4$ is reproduced below:
\be
M_Q
&=&\(\sum_{i=1}^4k_i^2\)\({b_{12}b_{34}\over s}\+{b_{41}b_{23}\over t}\)
-\[
{b_{12}\over s}\(d_3c_{43}+d_4c_{34}\)+{b_{41}\over t}\(d_2c_{32}+d_3c_{23}\)\right.\nn\\
&+&\left.{b_{23}\over t}(d_1c_{41}+d_4c_{14})+{b_{34}\over s}\(d_1c_{21}\+d_2c_{12}\)\].\nn\ee
Let us use it to verify partial gauge invariance. Since $\d_2(d_2)=k_2^2$,
\be
\d_2(M_Q)&=&k_2^2\[\({c'_{12}b_{34}\over s}+{c'_{32}b_{41}\over t}\)-\({b_{41}c_{32}\over t}+{b_{34}c_{12}\over s}\)\]+\cdots\nn\\
&=&-\h k_2^2\[{d_1b_{34}\over s}+{d_3b_{41}\over t}\]+\cdots,
\nn\ee
where the ellipses represent terms without $k_2^2$. Thus the $k_2^2$ coefficient of $\d_2(M_Q)$
vanishes in the zeroth order of the off-shell parameters. Similarly, the $k_i^2$ coefficients of the other $\d_i(M_Q)$ also vanish in the zeroth order,
thereby verifying that $M_Q$ possesses partial gauge invariance. 

Next,  to illustrate the power of partial gauge invariance, we will use it to constrain 
the possible dependence  of $M_Q$. For simplicity, let us  assume 
the absence of $b_{13}$ and $b_{24}$.  
On dimensional grounds, each term of $M_Q$ must contain $\e_1,\e_2,\e_3,\e_4$ once and $k$ twice
in the numerator. The denominator could be either $s=s_{12}=s_{34}$ or $t=s_{41}=s_{23}$.
The numerator must also contain at least one off-shell parameter, therefore its allowed forms are
confined to $b_{ij}b_{kl}k_m^2$ and $b_{ij}c_{kp}d_l$, with $(ijkl)$ being a permutation of (1234). 

With $b_{13}$ and $b_{24}$  absent, $(ij)$ in these terms must be either (12) or (34). First consider the term $ b_{12}b_{34}k_m^2/s_{12}$. Since $M_Q$ is cyclic permutation invariant, $M_Q$ must consist of the combination
\be
&&\a\[{b_{12}b_{34}\over s_{12}}k_m^2+{b_{23}b_{41}\over s_{23}}k_{m\+1}^2+{b_{34}b_{12}\over s_{34}}k_{m\+2}^2\+{b_{41}b_{23}\over s_{41}}k_{m\+3}^2\]\nn\\
=&&\a\[{b_{12}b_{34}\over s}(k_m^2+k_{m\+2}^2)+{b_{23}b_{41}\over t}(k_{m\+1}^2\+k_{m\+3}^2)\].\labels{bb}\ee
Under $\d_i$, to leading order $b_{ij}$ turns into $c_{ji}$, so in order to have partial gauge invariance,
the $bcd$ terms in $M_Q$ must be the following if $m=1$ or 3:
\be
-{\a\over s}\[b_{34}c_{21}d_1+c_{43}b_{12}d_3\]-{\a\over t}\[b_{41}c_{32}d_2+b_{23}c_{14}d_4\].\nn\ee

Applying a similar argument to the case when $m=2$ or 4, and to the situations when the starting denominator is $t$
rather than $s$, we conclude that  $M_Q$ must be equal to
\be
M_Q&=&{\a_1\over s}\[b_{12}b_{34}(k_1^2+k_{3}^2)-b_{34}c_{21}d_1-c_{43}b_{12}d_3\]
+{\a_1\over t}\[b_{23}b_{41}(k_{2}^2\+k_{4}^2)-b_{41}c_{32}d_2-b_{23}c_{14}d_4\]\nn\\
&+&{\a_2\over s}\[b_{12}b_{34}(k_2^2+k_{4}^2)-b_{34}c_{21}d_1-c_{43}b_{12}d_3\]
+{\a_2\over t}\[b_{23}b_{41}(k_{1}^2\+k_{3}^2)-b_{41}c_{32}d_2-b_{23}c_{14}d_4\]\nn\\
&+&{\a_3\over t}\[b_{12}b_{34}(k_1^2+k_{3}^2)-b_{34}c_{21}d_1-c_{43}b_{12}d_3\]
+{\a_3\over s}\[b_{23}b_{41}(k_{2}^2\+k_{4}^2)-b_{41}c_{32}d_2-b_{23}c_{14}d_4\]\nn\\
&+&{\a_4\over t}\[b_{12}b_{34}(k_2^2+k_{4}^2)-b_{34}c_{21}d_1-c_{43}b_{12}d_3\]
+{\a_4\over s}\[b_{23}b_{41}(k_{1}^2\+k_{3}^2)-b_{41}c_{32}d_2-b_{23}c_{14}d_4\].\nn
\ee
The result agrees with \eq{mq4} if we set $\a_1=\a_2=1$ and $\a_3=\a_4=0$.

\section{$\boldsymbol{ n\ge 5}$ Amplitudes}

\subsection{Organization of Feynman Diagrams}
Amplitudes of large $n$ contain
 many Feynman diagrams, and each contains many terms.  These terms can be organized in the following way.
  
 A Feynman diagram without a four-gluon vertex contains $n$ polarization vectors, $(n\-2)$ triple-gluon vertices,
 and $(n\-3)$ propagators, giving rise to a numerator of the form 
 $b_{i_1i_2}b_{i_3i_4}\cdots b_{i_{2k\-1},i_{2k}}c'_{i_{2k\+1}
 j_{2k\+1}}\cdots c'_{i_nj_n}a'_{j_1j_2}\cdots a'_{j_{2k\-3}j_{2k\-2}}$, where $I=(i_1i_2\cdots i_n)$ is a permutation of $(12\cdots n)$. Terms
 with different $j_m$'s can mix through momentum conservation, but there is no way to combine terms with  different $k$ or different $I$, thus it is useful to group together terms with the same $k$ and $I$.
 A Feynman diagram
 contains terms with different $k$'s and $I$'s, but each of its sub-diagrams contains a fixed $k$ and a fixed $I$ 
 
If four gluon vertices are present, each vertex simply eliminates a propagator and a pair of $k$'s in the numerator.
 
 For on-shell amplitudes,  $M_F(a',b',c')=M_P(a',b',c')$. Instead of using Feynman rules and Feynman
 diagrams, the amplitude can also be computed using
 Pfaffian diagrams obtained from \eq{ms} \cite{LY3,PD}. Like the Feynman sub-diagrams, each Pfaffian diagram has
 a fixed $k$ and a unique $I$ structure, but unlike Feynman sub-diagrams, Pfaffian diagrams do not contain internal
 momenta, so the necessity of
expanding  internal momenta into sums of external momenta is avoided, thereby resulting in fewer terms at the end \cite{LY3, PD}.
 
 For off-shell amplitudes, the decomposition $M_F=M_P+M_Q$ again results in fewer terms. 
 $M_P$ can be computed using Pfaffian diagrams as before, simply by replacing $a'$ with $a$ and $c'$ with $c$. 
 The computation of $M_Q$
is relatively simple because many Feynman diagrams do not contribute to $M_Q$, and  for those that do only
some off-shell parameters appears.
Furthermore, partial gauge invariance can be used to check the calculation. Thus both on-shell and off-shell,
there is an advantage to use the CHY formalism to compute Yang-Mills amplitudes. It results in having
fewer terms at the end.

We now illustrate  the computation of part of  $M_Q$ for $n=5$, and how partial gauge invariance can be
used to check this calculation.

\subsection{$\boldsymbol{M_Q}$ for $\boldsymbol{n=5}$}
Fig.6 shows all the sub-diagrams that contribute to terms proportional to $b_{12}/s_{12}s_{45}$. When 
$d_i$ appears in a sub-diagram, its $i$ is surrounded by a square. When $k_i^2$ appears, its $i$ is surrounded
by a circle. For example, no line in sub-diagram (h) has a square or a circle, so that diagram
carries no off-shell parameter and does not contribute to $M_Q$. Lines 4 and 5 in (d) and (e) are not surrounded by a circle so $k_4^2$ and $k_5^2$ are not present in the $M_Q$ of these diagrams.

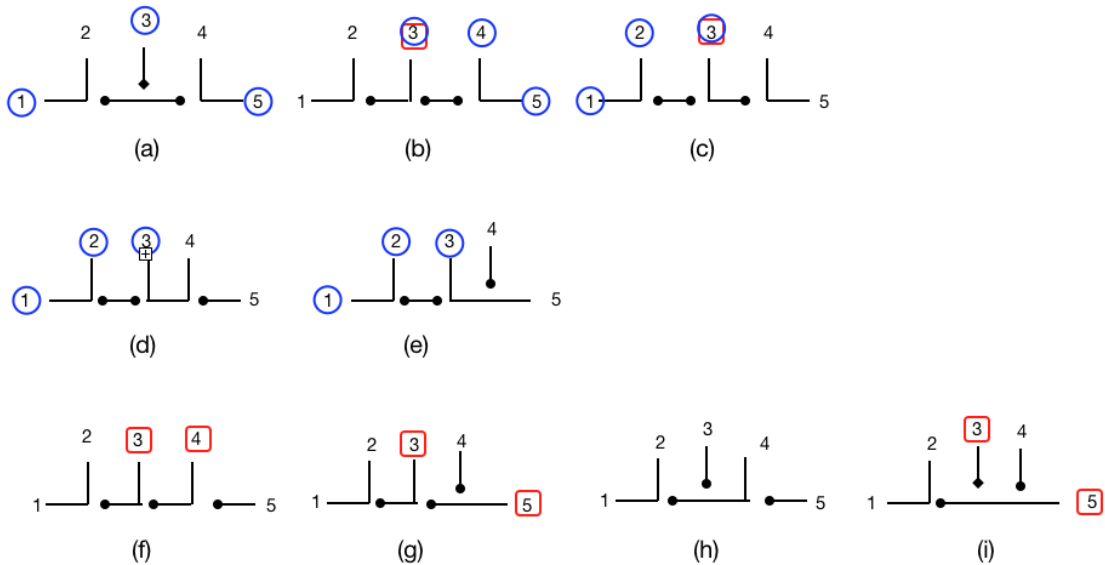
\bc\igw{15}{Fig6}\\ Fig.6\quad Sub-diagrams contributing to  $b_{12}/s_{12}s_{45}$
terms of $M_Q$ for $n=5$\ec

The contributions to $M_Q$ from
diagrams (a), (b), (c) are

\be
-\h b_{12}b_{45}&&\[(a_{13} \-2 a_{14} \-a_{23} \+2 a_{25} \+a_{34} \-a_{35} )\ d_{3}\right.\nn\\
&&+(\-6 c_{34}\-2 c_{35}\+d_{3})\  k_{1}^2+ (\-2 c_{34}\+2 c_{35}\+d_{3})\  k_{2}^2+ (4 c_{32}\+2 c_{34}\+6 c_{35})\  k_{3}^2\nn\\
&&\left. +(4 c_{32}\+2 c_{34}\+2 c_{35}\-d_{3})\  k_{4}^2+ (4 c_{32}\-2 c_{34}\-2 c_{35}\-d_{3})\  k_{5}^2\],
\labels{bbc5}\ee
and the contributions from diagrams (f), (g), (i) are
\be
\h b_{12} \[4(c_{54} c_{43}-c_{45}c_{53})\ d_{3}+ c_{54}(\-2c_{31}\+2 c_{32}\-d_{3})\  d_{4}+c_{45} (2 c_{31}\+6 c_{32}\-d_{3})\ d_{5}\]
\labels{bcc5}\ee
Let us use these expressions to verify partial gauge invariance, which demands $\d_i(M_Q)$ to contain no $k_i^2$ term in the zeroth order. This means
that after we make the replacements $b_{ij}\to c_{ji}, c_{ij}\to a_{ji}, d_i\to k_i^2$, the  coefficient of $k_i^2$
in $M_Q$ without any off-shell parameters must be identically zero. This is true for all $b_{ij}$ and all
propagators, so 
those terms proportional
to the same product of $b$ with the same propagator in $\d_i(Q)$ must be identically zero in the zeroth order
as well. 

The factor $b_{12}$ in Fig.6 will not
be altered by $\d_i(M_Q)$ only for $i=3,4,5$, so without including more diagrams, we can only verify partial gauge invariance from Fig.6 for $i=3,4,5$.  Diagrams 6(d) and 6(e) do not  contain $k_4^2$ and
$k_5^2$, so they can be ignored for the verification of $i=4$ and $i=5$. It is then easy to 
see from \eq{bbc5} and \eq{bcc5} that partial gauge invariance is indeed valid for these two $i$'s.

If we concentrate on terms of $M_Q$ proportional to $b_{12}b_{45}/s_{12}s_{45}$,
only diagrams 6(a), 6(b), 6(c) contribute and only \eq{bbc5} is relevant.
After applying $\d_3$ to it,   the leading coefficient  of $-\h k_3^2b_{12}b_{45}/s_{12}s_{45}$ is seen to be
\be
(a_{13} \-2 a_{14} \-a_{23} \+2 a_{25} \+a_{34} \-a_{35} )
+ (4 a_{23}\+2 a_{43}\+6 a_{53})=2(a_{12}\-a_{45})=s_{12}-s_{45},\nn\ee
where $\sum_{j\not=i}a_{ij}=0$ of \eq{conda}, and the relations $2a_{12}=s_{12}, 2a_{45}=s_{45}$, have been used. Since the propagator for this term is $1/s_{12}s_{45}$, the resulting numerator above cancels one
factor of the propagator leaving the coefficient of the double pole to be zero, so indeed leading coefficient
of $k_3^2b_{12}b_{45}/s_{12}s_{45}$ is indeed zero, as demanded by partial gauge invariance.

\section{Conclusion}
It is difficult for an S-matrix theory to incorporate interaction because it knows nothing about the local
space-time structure. An exception is the CHY theory, which with the guide of \M invariance, is
able to reproduce massless tree amplitudes for $\f^3$, Yang-Mills, gravity, and many other theories.
Whether it can replace the Lagrangian or not depends on whether off-shell amplitudes can also be made \M invariant,
and whether such invariant amplitudes can reproduce the correct tree amplitudes coming from a Lagrangian.
For scalar particles, it is known that the CHY formula can be modified so that off-shell amplitudes
remain \M invariant and reproduces the $\f^3$ interaction. For the Yang-Mills theory considered in this article,
it turns out that the CHY formula can also be modified to retain \M invariance for off-shell kinematics, but
the modified amplitude $M_P$ is not locally gauge invariant and therefore is not the correct Yang-Mills amplitude.
A complementary amplitude $M_Q$ must be added to restore local gauge invariance, but the construction of
this extra amplitude requires the Lagrangian or the off-shell Feynman diagrams, as  \M invariance provides
no clue.  Although neither
$M_P$ nor $M_Q$ is locally gauge invariant, both are partially gauge invariant, a useful property
that can be used to verify calculations and to simplify the Yang-Mills amplitude in the way discussed in the last section.

\end{document}